# Rhodium Doped Manganites : Ferromagnetism and Metallicity

B. Raveau, S. Hébert*, A. Maignan, R. Frésard and M. Hervieu

Laboratoire CRISMAT, ISMRA, 6 Boulevard du Maréchal Juin
14050 CAEN Cedex, FRANCE.

and

D. Khomskii

Solid State Physics Laboratory, Materials Science Centre
University of Groningen, Nijenborgh 4, 9747 A6 Groningen, The Netherlands

**Abstract**

The possibility to induce ferromagnetism and insulator to metal transitions in small A site cation manganites $Pr_{1-x}Ca_xMnO_3$ by rhodium doping is shown for the first time. Colossal magnetoresistance (CMR) properties are evidenced for a large compositional range ($0.35 \leq x < 0.60$). The ability of rhodium to induce such properties is compared to the results obtained by chromium and ruthenium doping. Models are proposed to explain this behavior.

*corresponding author

sylvie.hebert@ismra.fr

fax : 33 (0)2 31 95 16 00

# I. INTRODUCTION

The perovskite manganites $Ln_{1-x}Ca_xMnO_3$ often exhibit, at low temperature, a charge ordered state [1-6], whose metastable character is crucial for the appearance of colossal magnetoresistance (CMR). The CMR effect in these cases is indeed based on the competition between the insulating charge ordered state and the ferromagnetic metallic state, that takes place when a magnetic field is applied, leading also to the phase separation phenomena [7-13]. In those manganites, the stability of the charge ordered (CO) state increases as the size of the A-site cation decreases [14-17] so that the possibility to obtain a ferromagnetic metallic (FMM) state under a magnetic field is hindered for small A-site cations, and consequently CMR disappears.

An interesting route to induce CMR in such perovskites consists of doping Mn sites with foreign cations, since such a method weakens and even destroys the ordering of $Mn^{3+}$ and $Mn^{4+}$ species. Nevertheless, the collapse of CO is not sufficient to induce CMR, since it should also enhance ferromagnetism and metallicity. For this reason, only some magnetic cations, such as cobalt, nickel [18], chromium [19-20] and recently ruthenium [21-24], were found to be effective dopants. Rhodium, because of its two stable electronic configurations, $Rh(III)-d^6$ and $Rh(IV)-d^5$ is a potential dopant for CMR manganites. We report herein on the possibility to induce both ferromagnetism and metallicity in $Pr_{1-x}Ca_xMnO_3$ manganites with by doping with rhodium. We show that the behavior of rhodium is rather similar to chromium, but with smaller Curie temperatures ($T_C$) and larger resistivities.

# II. EXPERIMENTAL

The manganites $Ln_{1-x}Ca_xMn_{1-y}Rh_yO_3$ were synthesized from intimate mixtures of oxides $Pr_6O_{11}$ or $Sm_2O_3$, CaO, $Mn_2O_3$ and $Rh_2O_3$ first heated in air at 1050°C for 12 hrs, then sintered in the form of bars up to 1400°C for 12 hrs and slowly cooled down to room temperature at $100K.hr^{-1}$. The magnetic measurements were performed with a SQUID magnetometer under 1.45T, whereas the transport measurements were carried out with a Quantum Design physical properties measurements system (PPMS), under 0 and 7T (four-probe method). The purity of the samples was checked by electron diffraction (ED) using a 200kV microscope, and the cationic composition was determined by energy dispersive spectroscopy (EDS), using a Kevex analyzer. The analyses carried out for numerous crystallites show an homogeneous distribution of the cations and an actual composition very close to the nominal one, within the limit of the technique accuracy. T-dependent ED and lattice imaging were also carried out between 92K to 400K. Thermopower measurements were made under zero applied magnetic field using a four point steady state method with separate measuring and power contacts. More details of the experimental set-up are given in [25].

## III. RESULTS

The rhodium substitution for manganese in the hole doped manganites $Pr_{0.8}Ca_{0.2}MnO_3$ and $Pr_{0.7}Ca_{0.3}MnO_3$ does not influence significantly the magnetic and transport properties of these compounds: these oxides remain insulators and ferromagnetic whatever the rhodium level up to 10% Rh, with a $T_C$, taken at the inflection point, close to 120K, to be compared to 120K for the pristine compounds [26]. Similarly, the magnetic moment at 5K for $Pr_{0.8}Ca_{0.2}MnO_3$, M = 3.7$\mu_B$ is not greatly affected, leading to M = 3.2$\mu_B$ for $Pr_{0.8}Ca_{0.2}Mn_{0.9}Rh_{0.1}O_3$. Nevertheless, the doping with rhodium reinforces significantly ferromagnetism as the calcium content increases, as shown for $Pr_{0.7}Ca_{0.3}MnO_3$, which magnetic moment at 5K, M = 1.7$\mu_B$ is increased up to M = 2.7$\mu_B$ by doping with 5% Rh.

But the most spectacular effect appears for the manganites $Pr_{1-x}Ca_xMnO_3$ with $0.35 \leq x \leq 0.50$, for which the doping with rhodium induces a dramatic increase of ferromagnetism as illustrated by the M(T) curves of $Pr_{0.65}Ca_{0.35}Mn_{1-y}Rh_yO_3$ (Fig. 1a) and of $Pr_{0.5}Ca_{0.5}Mn_{1-y}Rh_yO_3$ (Fig. 1b) which show that the magnetic moment of the pristine oxide at 5K, smaller than 0.5$\mu_B$ is increased up to 3.2$\mu_B$ by doping with rhodium. As shown for $Pr_{0.5}Ca_{0.5}Mn_{1-y}Rh_yO_3$, the magnetic moment generally increases regularly with the rhodium content reaching the magnitude close to the theoretical value (~ 3.5$\mu_B$) for y = 0.06 and then decreases again as y increases (Fig. 1b). Concomitantly, the magnetization maximum at about 250K, which is characteristic of the charge-ordering setting in $Pr_{0.5}Ca_{0.5}MnO_3$, tends to disappear as the Rh content increases and is no more observed for y = 0.06 (Inset of Fig. 1b). Note also that $T_C$ does not vary dramatically, being close to 125K for y = 0.06 and decreasing down to 80K for y = 0.10. Correspondingly, the $\rho$(T) curves (Fig. 2) are characterized by a maximum when doped with rhodium, indicating a tendency to a transition from an insulating paramagnetic state to a ferromagnetic metal like one, at decreasing temperature. Nevertheless, the value of the resistivity at low temperature remains high (> 5.10$^{-2}\Omega$.cm), so that most often the samples cannot be described even as bad metals, contrary to what is observed for chromium or ruthenium doping [19-24]. This is especially the case of the limit compounds $Pr_{0.65}Ca_{0.35}Mn_{1-y}Rh_yO_3$ (Fig. 2a) for which a bump in the resistivity is observed around 50K for y = 0.05-0.10, but the resistivity at low temperature, i.e. at 5K, is still very high ($> 10^3\Omega$.cm), though the $\rho$ value at 5K is too high to be measured in the undoped phase [26]. For larger calcium contents, the metallicity is significantly increased by Rh doping as shown for the $Pr_{0.5}Ca_{0.5}Mn_{1-y}Rh_yO_3$ series (Fig. 2b). For the latter series, one indeed observes a peak shaped $\rho(T)$ curve whatever the Rh content, ranging from 2% to 10%, and the resistivity is decreased by several orders of magnitude with respect to the undoped material. The resistivity at low temperature (5K) decreases indeed from $\geq 10^6\Omega$.cm for y = 0 to $7.10^{-2}\Omega$.cm for y = 0.04 and increases again with y, reaching $30\Omega$.cm for y = 0.10. Thus a transition from an insulating to a semi-metallic or poorly metallic state is clearly observed for 4%-5% Rh. Note also that the transition temperature ($T_{peak} \sim 110K$) follows the Curie temperature (Fig. 1b), increasing as y increases up to 4%, and then decreasing as y increases from 4% to 10%. These results will be discussed below in the picture of phase separation [11].

For the electron rich region, the effect of Rh-doping drops abruptly, as shown for $Pr_{0.4}Ca_{0.6}Mn_{1-y}Rh_yO_3$ (Fig. 3) which exhibits a much lower magnetic moment ranging from $0.25\mu_B$ for y = 0.05 to $0.55\mu_B$ for y = 0.10 (Fig. 3a) and remains insulating (Fig. 3b) whatever the rhodium content. Note that for $Pr_{0.2}Ca_{0.8}Mn_{1-y}Rh_yO_3$, practically no ferromagnetism is induced by Rh-doping ($M < 0.1\mu_B$) and the oxides remain insulating similarly to the pristine y = 0 composition [16].

As previously observed for the doping with magnetic cations [20], the effect of Rh-doping decreases with the average size of the A-site cation. For instance, the doping of $Sm_{0.5}Ca_{0.5}MnO_3$ with rhodium induces only small ferromagnetic components, $M = 0.25\mu_B$ and $M = 0.85\mu_B$ (Fig. 4a) for y = 0.05 and 0.10 respectively, and, moreover, the materials remain insulating (Fig. 4b).

These results show that the doping of $Pr_{1-x}Ca_xMnO_3$ manganites with rhodium induces ferromagnetism and metallicity in a rather similar way as was found for chromium doping [19-20]. The most important effect of rhodium concerns its ability to induce ferromagnetism in the $Pr_{1-x}Ca_xMnO_3$ series with x ranging from 0.35 to 0.50, with magnetic moments very similar to those observed for Cr-doping. Nevertheless, the Curie temperatures of the Rh-doped manganites are smaller than those of the Cr-doped compounds reaching maximum values of 120K against 150K for the Cr doped ones [19-20], and the corresponding M(T) curves are generally smoother, indicating that Rh is less effective to induce ferromagnetism than Cr. This viewpoint is supported by the Rh-doped $Sm_{0.5}Ca_{0.5}MnO_3$ compounds which exhibit maximum magnetic moments at 5K of 0.25-0.85$\mu_B$ only, to be compared to the Cr-doped ones which reach maximum values of 1.2-2.0$\mu_B$ [20]. The less effective ability of Rh to induce ferromagnetism compared to Cr, is demonstrated by comparing the series $Pr_{0.4}Ca_{0.6}Mn_{1-y}Rh_yO_3$ which exhibit magnetic moments at 5K of only 0.55$\mu_B$, with the series $Pr_{0.4}Ca_{0.6}Mn_{1-y}Cr_yO_3$ for which magnetic moments of 2.5$\mu_B$ are observed [20]. Finally Rh-doping differs fundamentally from Cr-doping by its significantly smaller efficiency to induce metallicity: we indeed observe for instance that the resistivity of the Rh-doped $Pr_{0.5}Ca_{0.5}MnO_3$ compounds is more than one order of magnitude larger than the corresponding Cr-doped samples.

The ability of rhodium to induce both ferromagnetism and a certain tendency to metallicity, makes that colossal magnetoresistance (CMR) can be expected for the Rh-doped manganites. This is indeed observed for the $Pr_{0.65}Ca_{0.35}Mn_{1-y}Rh_yO_3$ series (Fig. 5a) which exhibits the highest resistivity ratios under 7T at 50K ($\rho_{0T}/\rho_{7T}$ reaches $10^7$ at 50K) and for the $Pr_{0.5}Ca_{0.5}Mn_{1-y}Rh_yO_3$ series (Fig. 5b) which is characterized by a maximum value of the resistance ratio at $T_C$ ($\rho_{0T}/\rho_{7T} \approx 10^2$). Note that the series $Pr_{0.4}Ca_{0.6}Mn_{1-y}Rh_yO_3$, exhibits CMR properties with resistivity ratios close to $10^5$ at 50K (Fig. 5c) though its ferromagnetic component is rather weak which indicates that these compositions lie close to the percolation threshold.

The Rh ability to progressively hinder the CO process at the benefit of the ferromagnetic metallic state as its content increases has been checked by electron diffraction as a function of temperature for $Pr_{0.5}Ca_{0.5}Mn_{0.95}Rh_{0.05}O_3$. At 92K, the sample is made of the coexistence of small CO regions with "non ordered" ones. The latter are characterized by an orthorhombic structure (Pbnma space group) with "$a_p\sqrt{2}$ x $2a_p$ x $a_p\sqrt{2}$" cell parameters where $a_p$ is the primitive perovskite cell parameter. A typical ED pattern of such a "non ordered" area is given in Fig. 6a ; it is similar to the room temperature one. In contrast, the CO regions observed at 92K exhibit ED patterns with extra peaks in incommensurate positions (Fig. 6b). This system of satellites is characteristic of the CO state, with a modulation vector qa*, so that a subcell "$\frac{1}{q}a_p\sqrt{2}$ x $2a_p$ x $a_p\sqrt{2}$" is obtained with an average q value close to 0.46. Note that the coexistence of ordered and "non ordered" domains was also observed in the $Pr_{0.5}Ca_{0.5}Mn_{0.95}Cr_{0.05}O_3$ [27], but with a lower modulation vector (q = 0.39). The lattice images recorded at 92K confirm that the sample is made of CO domains, a few tens nanometers wide, in a non ordered Pnma-type matrix. One example is given in Fig. 6d. The contrast consists of bright and less bright fringes. In the small CO areas, the local periodicity is $\mathbf{n}a_p\sqrt{2}$, with $\mathbf{n}$ = 2 and 3 (see in the middle part of the image, between small black arrows). In the rest of the grain, the distance between two fringes of equal intensity is $a_p\sqrt{2}$, ( $\mathbf{n}$ = 1), in agreement with the cell parameter of the non ordered structure (lower part of the image).

By warming the samples, the q value remains roughly constant in the range 92K-120K (Fig. 6e). Above 125K, the intensity of the satellites abruptly decreases whereas, concomitantly the q value smoothly decreases. This is illustrated in Fig.6c, recorded at 165K (same area as Fig. 6b). At 200K, the q value is close to 0.4 and the satellites are scarcely visible. Between 200K and 230K, there is no nodes but only diffuse streaks along a*, and above 230K they are no longer detectable. These results indicate that $T_{CO}$ is depressed in the doped $Pr_{0.5}Ca_{0.5}Mn_{0.95}Rh_{0.05}O_3$, in comparison with $T_{CO}$ = 250K for the pristine $Pr_{0.5}Ca_{0.5}MnO_3$ compound.

Accordingly, the field dependent magnetization M(H) curve registered at 5K for y = 0.05 exhibits a ferromagnetic behavior with a saturation at $3\mu_B$/f.u. i.e. close to the expected moment, (~ $3.5\mu_B$), which is very different from the low M values obtained for the CO-AFM y = 0.00 sample (Fig. 1b).

Another proof of the gradual disappearance of charge ordering induced by Rh doping is given by thermopower measurements. These measurements were performed on the sample $Pr_{0.5}Ca_{0.5}Mn_{0.95}Rh_{0.05}O_3$. Fig. 7 presents the results obtained as the sample was cooled from 320K down to 5K under zero field. They are compared to the results obtained for the undoped sample $Pr_{0.5}Ca_{0.5}MnO_3$ (y = 0). As previously described in [28], the charge ordering transition in $Pr_{0.5}Ca_{0.5}MnO_3$ at $T_{CO} \approx 250K$ is associated with a steep decrease of the thermopower leading to large negative values at low temperatures. Doping the compound with 5% of Rh strongly modifies thermopower. The steep decrease of S(T) is changed into a slowly decreasing value of S as T decreases. At $T \approx 90K$, as the sample enters the ferromagnetic and 'metallic' state as shown in Fig. 1b and 2b, the absolute value of thermopower goes back to very small values, characteristic of a 'metal'. This strong decrease of the absolute value of S in the high temperature insulating phase of the doped compound ($90K < T < T_{CO}$) shows that long-range ferromagnetic ordering has been efficiently stabilised by Rh doping.

In order to explain the role of rhodium in the magnetic and transport properties of these doped manganites, there are several aspects to consider. They include the electronic configuration of rhodium and the coupling to its environment. Also one needs to figure out why a (bad) metallic behavior sets in and the reason for the observed ferromagnetism. We now proceed to these considerations. Concerning first the electronic configurations of rhodium, we may consider, according to previous studies [29], that rhodium is either trivalent or tetravalent. The substitution of $Rh^{3+}$ for $Mn^{3+}$ in those oxides would not introduce any ferromagnetism, nor metallicity since the $(t_{2g})^6 e_g^0$ configuration of this species would lead to S = 0 as shown previously for d° cations ($Mg^{2+}$, $Al^{3+}$, $Ti^{4+}$, $Nb^{5+}$) [30]. The doping of these oxides with $Rh^{4+}$, with the $(t_{2g})^5$ configurations is on the other hand susceptible to introduce ferromagnetism and metallicity similar to $Ru^{4+}$ which is in the $(t_{2g})^4$ configuration [21-24]. Let us indeed consider the $Rh^{4+}$ doping of the charge ordered CE-type structure of $Pr_{0.5}Ca_{0.5}MnO_3$ (see projected 2D structure Fig. 8a), in an otherwise undisturbed charge-ordered state, in a static picture. This structure consists of FM zig-zag chains of $Mn^{3+}/Mn^{4+}$ cations, the coupling between chains being antiferromagnetic. In other words, one Mn atom is primarily coupled to two of its Mn neighbors through $e_g$ electrons, and weakly coupled to the other four through $t_{2g}$ electrons. One should first notice that the ionic radius of $Rh^{4+}$ (0.6Å) is substantially larger than that of $Mn^{4+}$ (0.53Å), but similar to $Mn^{3+}$ (0.64Å). Therefore the $Rh^{4+}$ ions should be sitting at the center of the $MnO_6$ octahedron in contrast to the off-centered location of $Mn^{4+}$ [31]. As a result, one expects that the coupling of $Rh^{4+}$ to its neighbors is going to be more symmetrical than it is the case for $Mn^{4+}$, thus leading to a local suppression of the charge ordering of the CE-type structure . As the rhodium content increases, the non-ordered regions will percolate, yielding a metallic behavior.

Having explained the role of rhodium substitution in establishing a metallic behavior, we turn to the magnetism. In this context one may expect that the larger $Rh^{4+}$ would "push away" surrounding oxygen as a result of which the orbitals of $Mn^{3+}$ surrounding $Rh^{4+}$ will all be directed away from it (see Fig. 8b). One immediately sees that, according to Goodenough-Kanamori rules, the spins of these flipped $Mn^{3+}$ ions would be reversed, creating small ferromagnetic clusters and establishing conducting "bridges" between neighboring parallel zigzag chains which, before introducing $Rh^{4+}$, were separated by the anti-parallel ones. This may be a factor contributing to the establishing of the FM state by rhodium doping.

There exists however yet another, and probably more important mechanism. Notably, there may occur valence fluctuations of the type $Mn^{3+} + Rh^{4+} \leftrightarrow Mn^{4+} + Rh^{3+}$. Note that these valence fluctuations would most probably occur via hopping of $e_g$ electrons, as a result of which the $Rh^{3+}$ ($d^6$) will be in an excited intermediate spin state ($t_{2g}^5 e_g^1$), similar to such state in $Co^{3+}$ [32]). They can effectively destroy charge ordering and provide an efficient mechanism for the ferromagnetism and metallic conductivity similar to the conventional double-exchange (this mechanism is analogous to the one responsible for the FM behavior of the double perovskites $Sr_2FeMoO_6$ [33]). This process is efficient for underdoped systems, when we have many $Mn^{3+}$ ions. However for over doped system ($x \geq 0.6$), $Rh^{4+}$ is predominantly surrounded by $Mn^{4+}$ ions, so that this mechanism becomes less and less efficient, in contrast to Ru doping where similar valence fluctuations are $Mn^{4+} + Ru^{4+} \leftrightarrow Mn^{3+} + Ru^{5+}$.

## IV. CONCLUSION

A novel anomalous ferromagnetic phase induced by rhodium substitution in charge-ordered manganites $Pr_{1-x}Ca_xMnO_3$ has been discovered. It extends from x = 0.35 to x = 0.5. With Curie temperatures around 100K it is found that Rh doping is less efficient than Cr or Ru doping. Nevertheless, the induced phase separation is responsible for the observed CMR properties. These properties are understood as following from the collapse of the charge ordering transition as probed by the dramatic change in the thermopower.

We gratefully acknowledge C. Martin for interesting discussions.

**References**

1. P. H. Woodward, D. E. Cox, T. Vogt, C. N. R. Rao and A. K. Cheetham, Chem. Mater. **11**, 3528 (1999).

2. C. H. Shen and S. W. Cheong, Phys. Rev. Lett. **76**, 4042 (1996).

3. F. Damay, Z. Jirak, M. Hervieu, C. Martin, A. Maignan, B. Raveau, G. André and F. Bourée, J. Magn. Magn. Mater. **190**, 221 (1998).

4. M. Hervieu, A. Barnabé, C. Martin, A. Maignan, F. Damay and B. Raveau, Eur. Phys. J. B **8**, 31 (1999), J. Mater. Chem. **8**, 1405 (1998).

5. Z. Jirak, Krupicka, Z. Simsa, M. Dlouha and S. Vratislav, J. Magn. Magn. Mater. **53**, 153 (1985).

6. C. H. Chen, S. W. Cheong and H. Y. Hwang, J. Appl. Phys. **81**, 4326 (1997).

7. P. Schiffer, A. P. Ramirez, W. Bao and S. W. Cheong, Phys. Rev. Lett. **75**, 3336 (1995).

8. P. G. Radaelli, D. E. Cox, M. Marezio, S. W. Cheong, P. Schiffer, and A. P. Ramirez, Phys. Rev. Lett., **75**, 4488 (1995).

9. G. Allodi, R. De Renzi, G. Gvidi, F. Licci and M. W. Piepper, Phys. Rev. B **56**, 6036 (1997).

10. A. Moreo, S. Yunoki and E. Dagotto, Science **283**, 2034 (1999).

11. D. Khomskii, Physica B **280**, 325 (2000).

12. M. Yu Kagan, K. I. Kugel and D. Khomskii, Cond. Mat. 0001245.

13. G. Allodi, R. De Renzi, F. Licciand and M. W. Piepper, Phys. Rev. Lett. **81**, 4736 (1998).

14. N. Kumar and C. N. R. Rao, J. Solid State Chem. **129**, 363 (1997).

15. A. Barnabé, M. Hervieu, C. Martin, A. Maignan and B. Raveau, J. Appl. Phys. **84**, 5506 (1998).

16. C. Martin, A. Maignan, M. Hervieu and B. Raveau, Phys. Rev. B **60**, 12191 (2000).

17. F. Damay, C. Martin, A. Maignan, M. Hervieu, Z. Jirak, G. André and F. Bourée, Chem. Mater. **11**, 536 (1999).

18. A. Maignan, F. Damay, C. Martin and B. Raveau, Mater. Res. Bull. **32**, 965 (1997).

19. B. Raveau, A. Maignan and C. Martin, J. Solid State Chem. **130**, 162 (1997).

20. A. Barnabé, A. Maignan, M. Hervieu, C. Martin and B. Raveau, Appl. Phys. Lett. **71**, 3907 (1997); Eur. Phys. J. B **1**, 145 (1998).

21. P. V. Vanitha, A. Arulraj, A. R. Raju and C. N. R. Rao, C. R. Acad. Sci. **IIc**, 595 (1999).

22. B. Raveau, A. Maignan, C. Martin, R. Mahendiran and M. Hervieu, J. Solid State Chem. **151**, 330 (2000).

23. C. Martin, A. Maignan, M. Hervieu, B. Raveau and J. Hejtmanek, Eur. J. Phys. **16**, 469 (2000).

24. C. Martin, A. Maignan, M. Hervieu, C. Autret, B. Raveau and D. I. Khomskii, to appear in Phys. Rev. B.

25. A. Maignan, C. Martin, M. Hervieu, B. Raveau and J. Hejtmanek, J. Appl. Phys, **89**, 2232 (2001).

26. Y. Tomioka, A. Asamitsu, H. Kuwahara, Y. Moritomo and Y. Tokura, Phys. Rev. B **53**, R1689 (1996).

27. M. Hervieu, C. Martin, A. Barnabé, A. Maignan, R. Mahendiran and V. Hardy. Solid State Science, at press.

28. J. Hejtmanek, Z. Jirak, Z. Arnold, M. Marysko, S. Krupicka, C. Martin, and F. Damay, J. Appl. Phys. **83**, 7204 (1998).

29. D. Y. Jung and G. Demazeau, Solid State Comm. **94**, 963 (1995); M. Zakhour, PhD thesis, University of Bordeaux I (July 2000).

30. F. Damay, C. Martin, A. Maignan and B. Raveau, J. Magn. Magn. Mater. **183**, 143 (1998).

31. Z. Jirak, F. Damay, M. Hervieu, C. Martin, B. Raveau, G. André and F. Bourée, Phys. Rev. B **61**, 1181 (2000).

32. M. A. Senaris-Rodriguez, and J. B. Goodenough, J. Solid State Chem. **118**, 323 (1995).

33. K. I. Kobayashi, T. Kimura, H. Sawada, K. Terakura, and Y. Tokura, Nature **395**, 677 (1998); D. Sarma et al., Phys. Rev. Lett. **85**, 2549 (2000).

**Figure captions**

Figure 1: T dependence of the magnetization M collected in 1.45T after a zero field cooled process for the series:

(a) $Pr_{0.65}Ca_{0.35}Mn_{1-y}Rh_yO_3$ and (b) $Pr_{0.5}Ca_{0.5}Mn_{1-y}Rh_yO_3$. y values are labelled in the graph. Inset of (b): enlargement in the vicinity of $T_{CO}$.

Figure 2: T dependence of the resistivity ρ registered upon cooling from 400K in the absence of magnetic field for the series:

(a) $Pr_{0.65}Ca_{0.35}Mn_{1-y}Rh_yO_3$ and (b) $Pr_{0.5}Ca_{0.5}Mn_{1-y}Rh_yO_3$.

Figure 3: $Pr_{0.4}Ca_{0.6}Mn_{1-y}Rh_yO_3$ : (a) M(T) and (b) ρ(T) curves.

Figure 4: $Sm_{0.5}Ca_{0.5}Mn_{1-y}Rh_yO_3$.: (a) M(T) and (b) ρ(T) curves.

Figure 5: T dependent resistivity ratios $\rho_0/\rho_{7T}$ obtained by dividing the resistivity in the absence of magnetic field ($\rho_0$) by the resistivity measured in the 7T($\rho_{7T}$). Both sets of data, $\rho_0$(T) and $\rho_{7T}$(T) are collected in cooling mode.

(a) $Pr_{0.65}Ca_{0.35}Mn_{1-y}Rh_yO_3$; (b) $Pr_{0.5}Ca_{0.5}Mn_{1-y}Rh_yO_3$; (c) $Pr_{0.4}Ca_{0.6}Mn_{1-y}Rh_yO_3$.

Figure 6: [010] orientation of the $Pr_{0.5}Ca_{0.5}Mn_{0.95}Rh_{0.05}O_3$ crystals.

a) ED pattern at 92K for a non ordered area and b) ED pattern at 92K for a CO area.

c) ED pattern at 165K for of the same CO area.

d) [010] lattice image recorded at 92K. In the CO areas the fringe spacing is $\mathbf{n}a_p\sqrt{2}$ ,with $\mathbf{n}$ = 2 and 3 (black numbering) and $a_p\sqrt{2}$ in the Pnma-type areas.

(e) T dependence of the modulation vector q for $Pr_{0.5}Ca_{0.5}MnO_3$ and for the remaining CO regions of $Pr_{0.5}Ca_{0.5}Mn_{0.95}Rh_{0.05}O_3$. The size of the symbols reflects the intensity of the satellites.

Figure 7: T dependence of the thermopower S for the compounds $Pr_{0.5}Ca_{0.5}MnO_3$ (y = 0) and $Pr_{0.5}Ca_{0.5}Mn_{0.95}Rh_{0.05}O_3$ (y = 0.05).

Figure 8: a) Orbital, charge and spin ordering of the CE-type structure.

b) Creation of ferromagnetic regions by $Rh^{4+}$ doping.

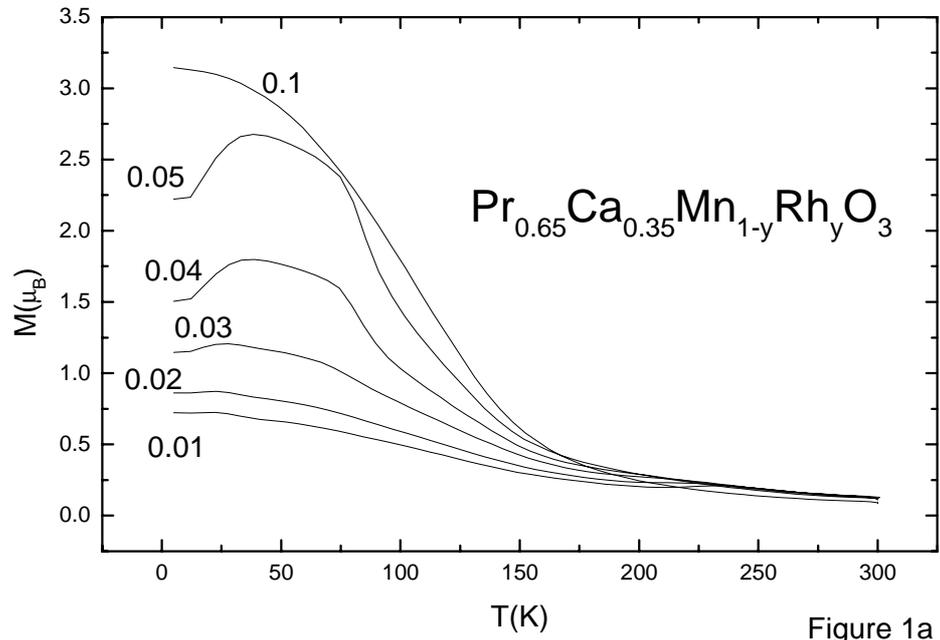

Figure 1a

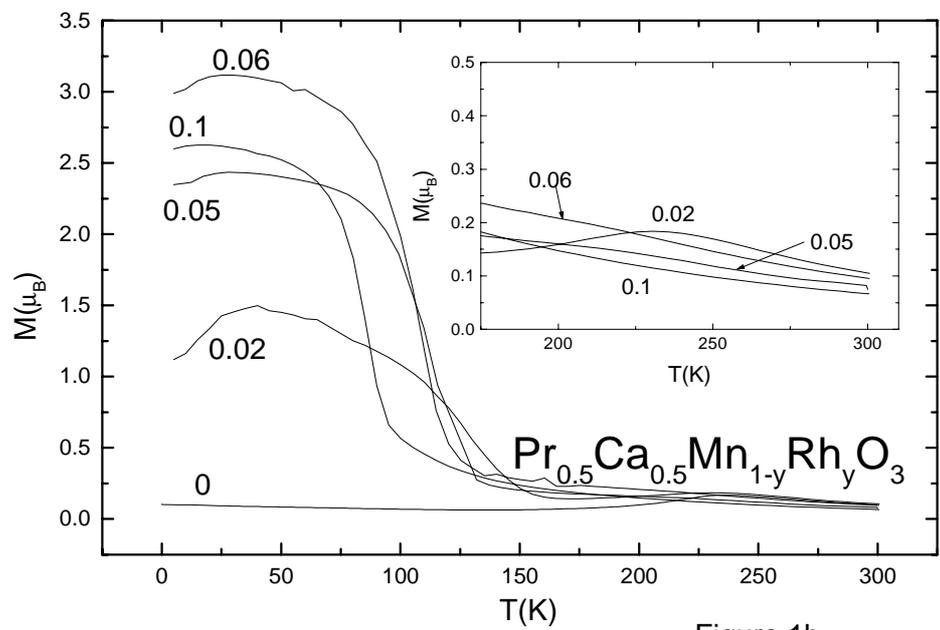

Figure 1b

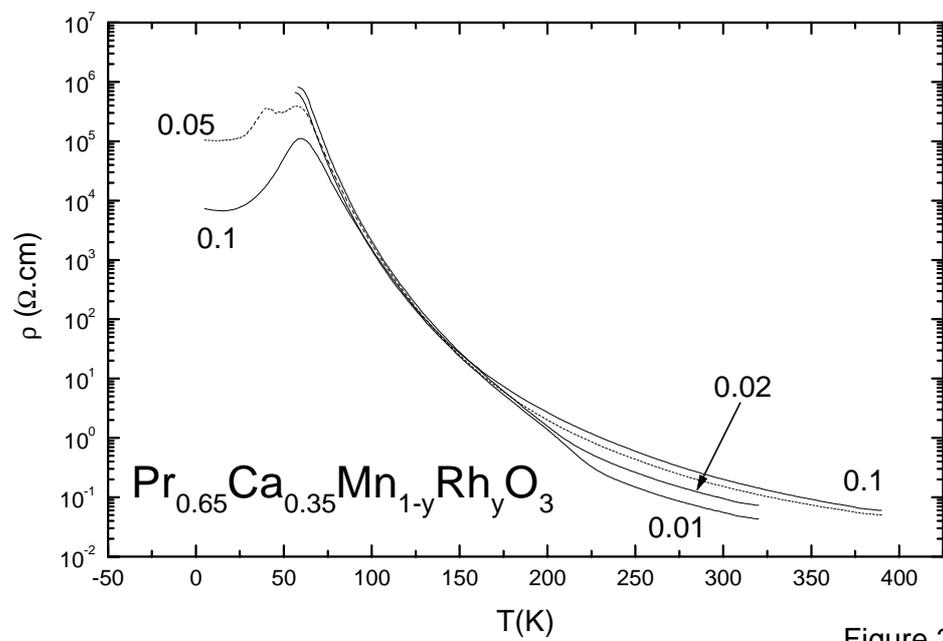

Figure 2a

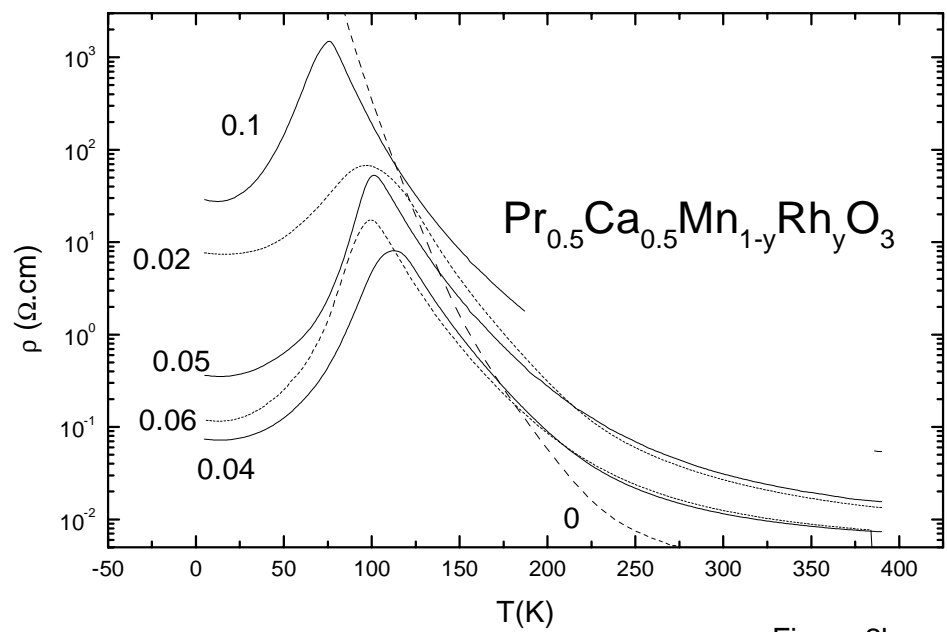

Figure 2b

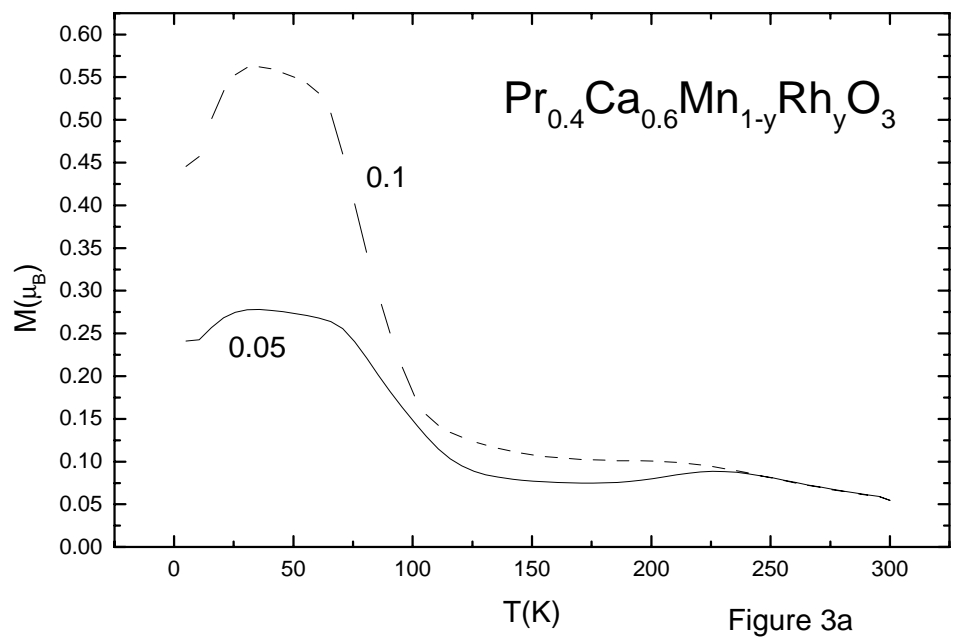

Figure 3a

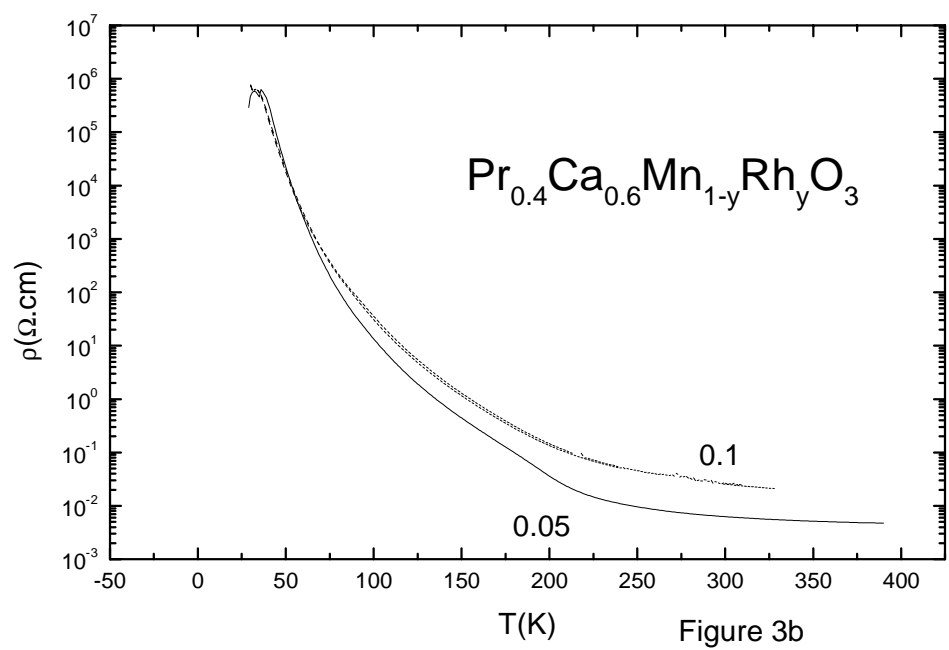

Figure 3b

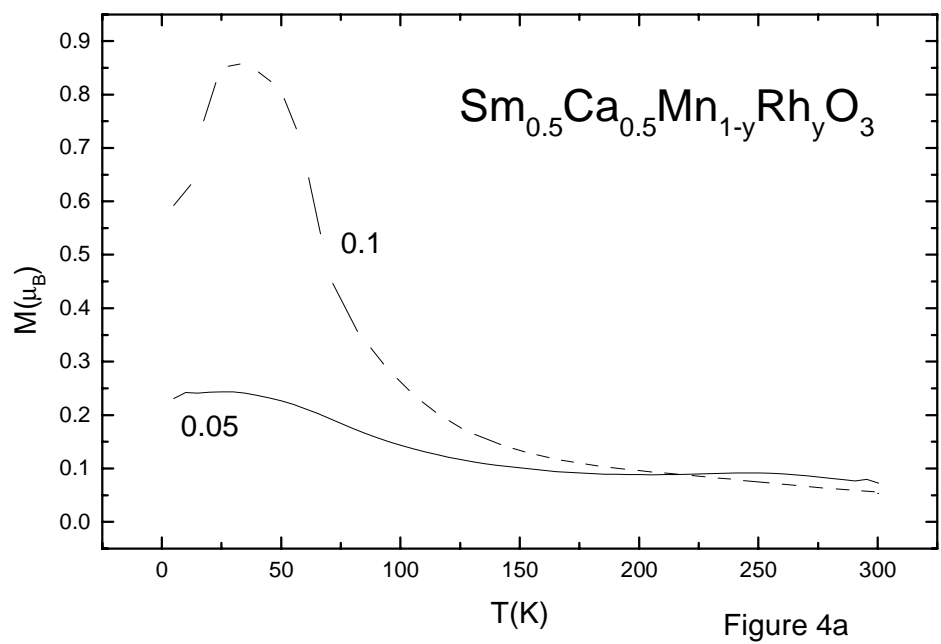

$Sm_{0.5}Ca_{0.5}Mn_{1-y}Rh_yO_3$

0.1

0.05

Figure 4a

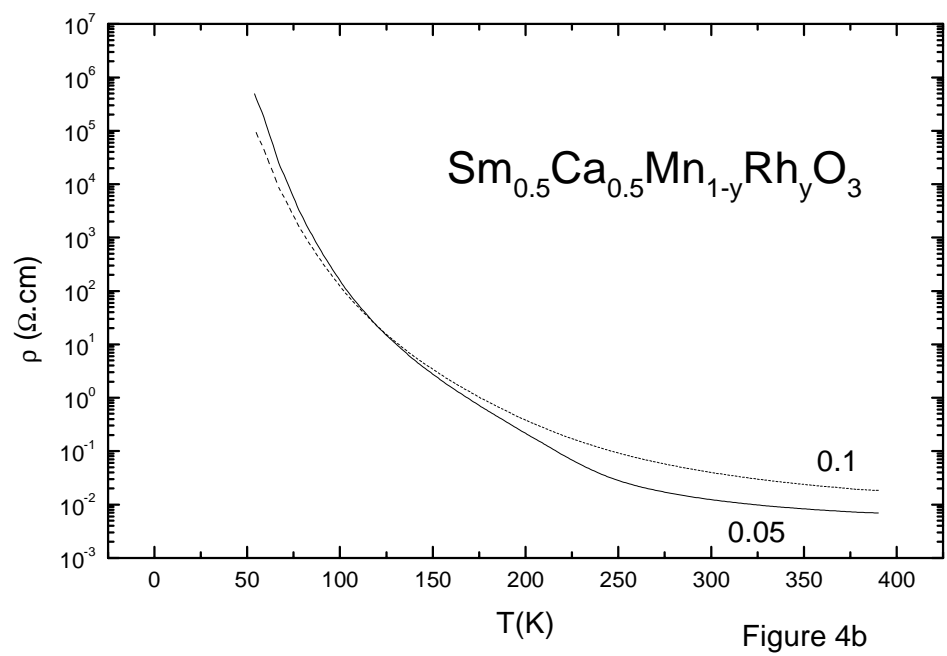

$Sm_{0.5}Ca_{0.5}Mn_{1-y}Rh_yO_3$

0.1

0.05

Figure 4b

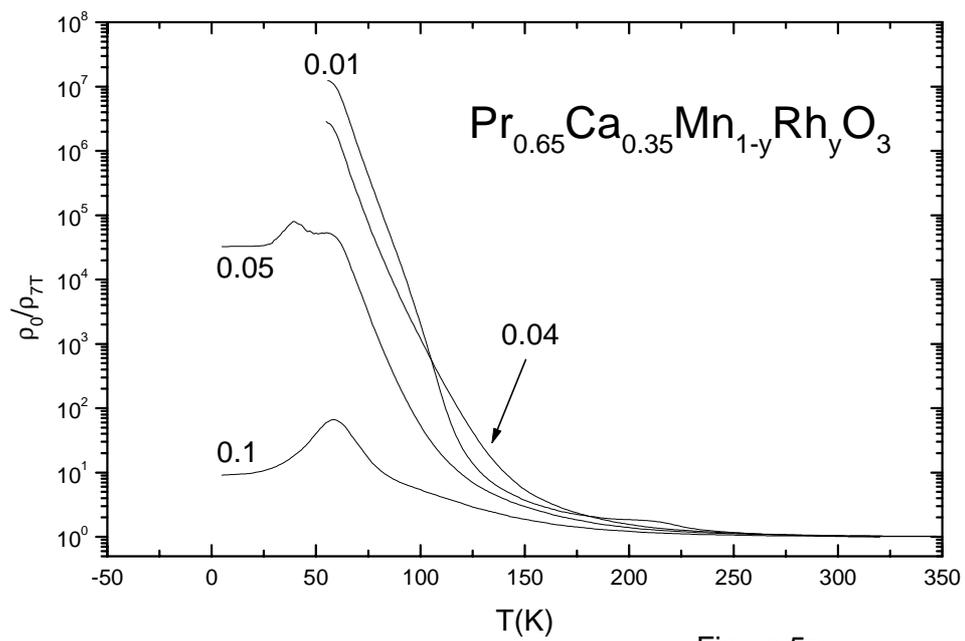

Figure 5a

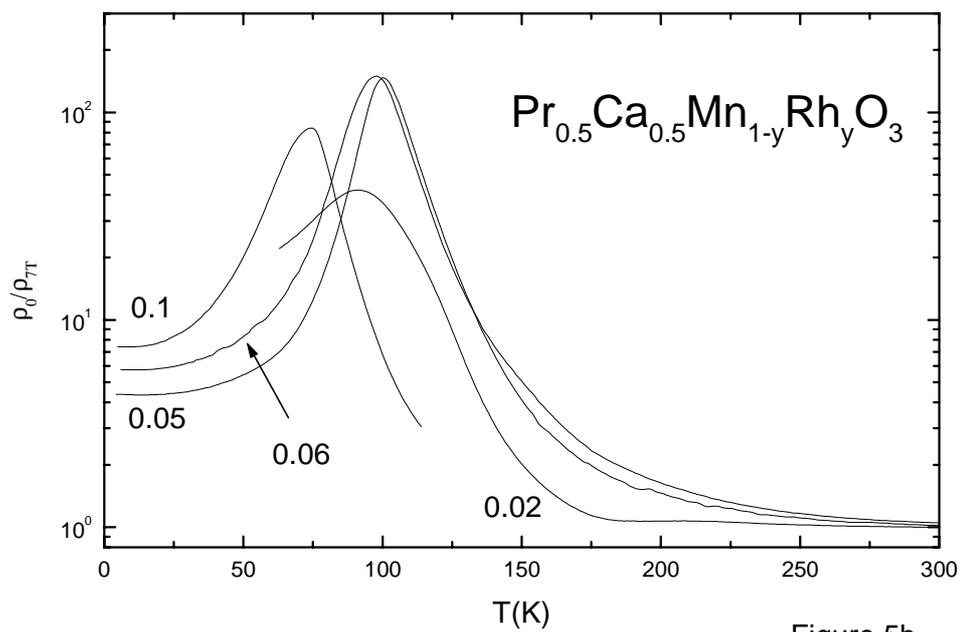

Figure 5b

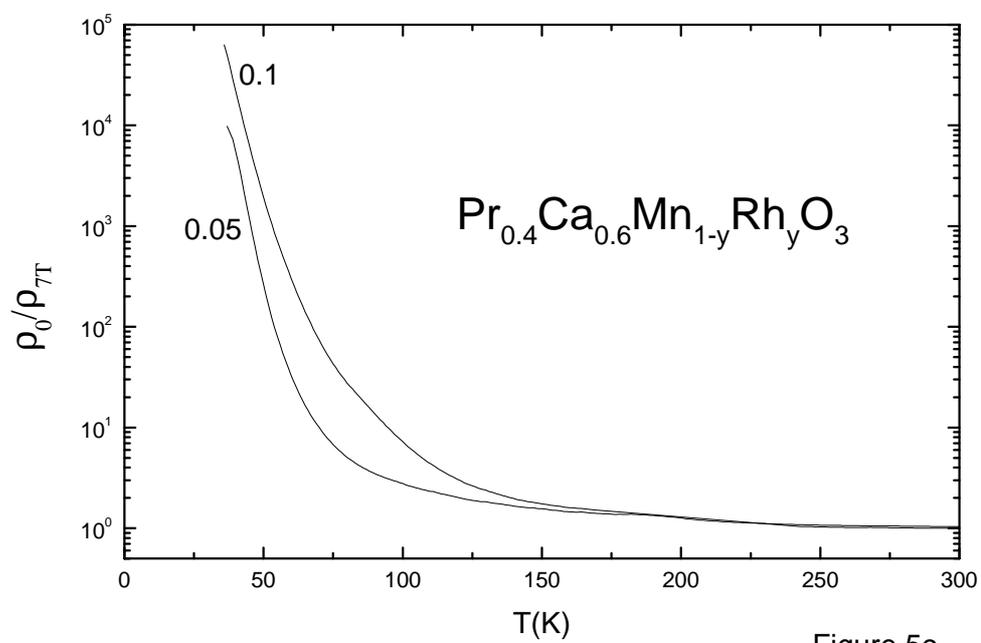

Figure 5c

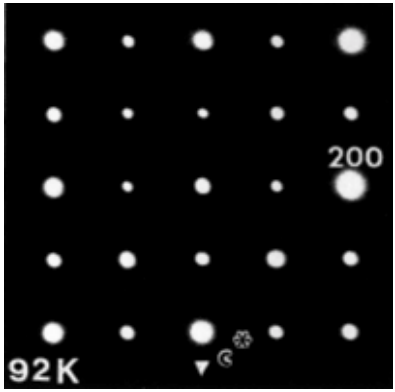

Figure 6a

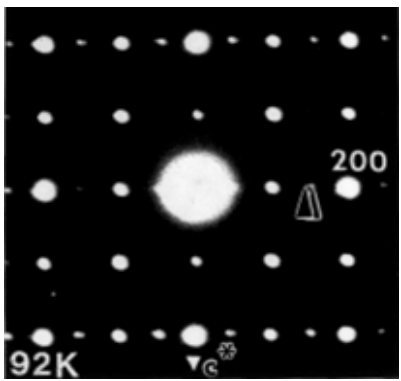

Figure 6b

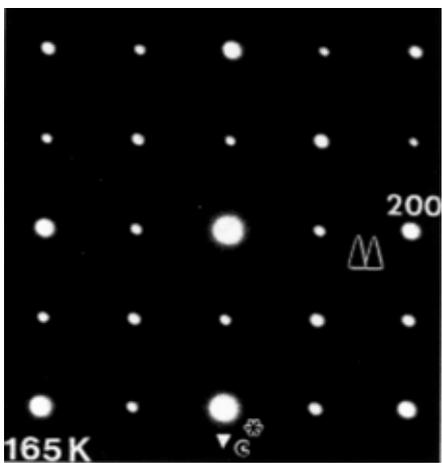

Figure 6c

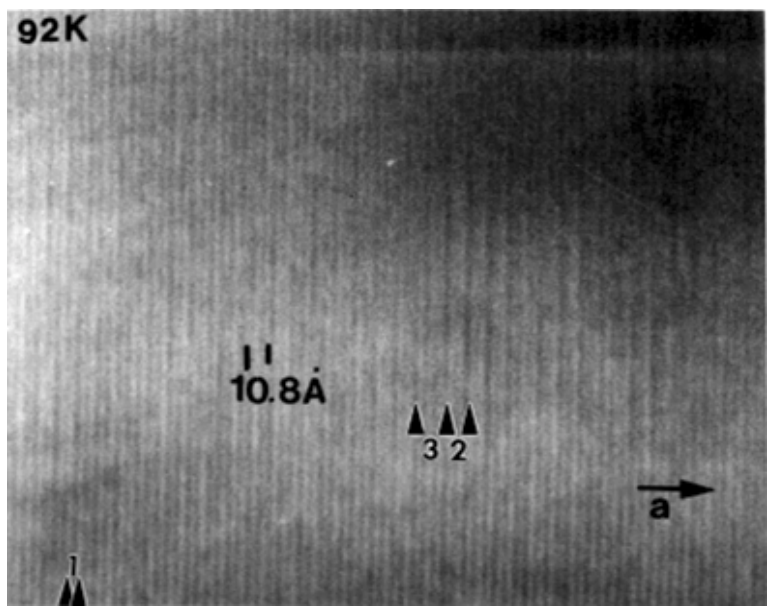

Figure 6d

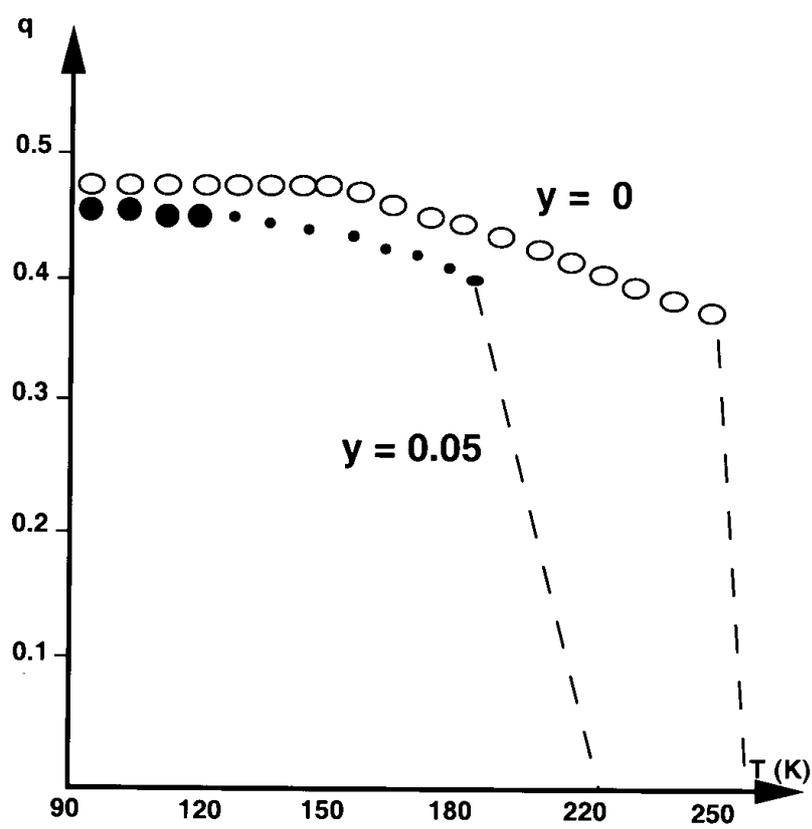

Figure 6e

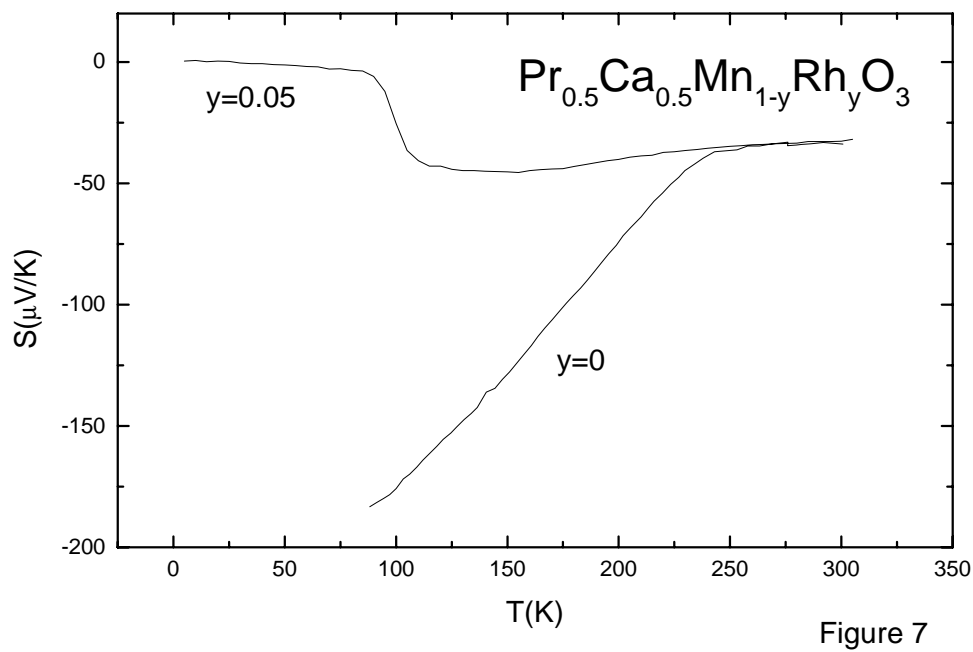

$Pr_{0.5}Ca_{0.5}Mn_{1-y}Rh_yO_3$

Figure 7

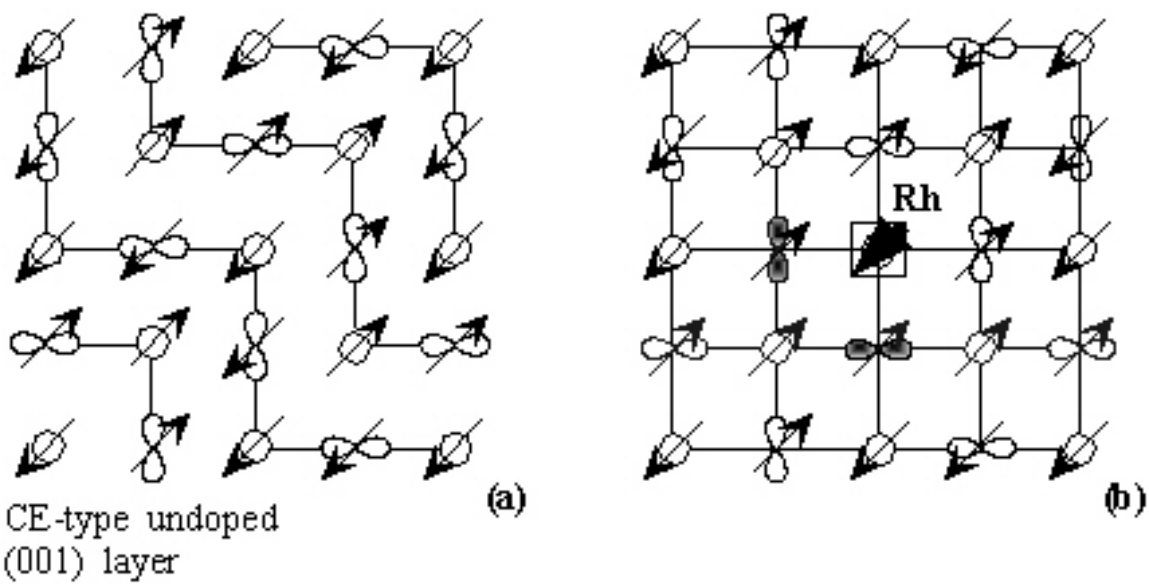

CE-type undoped
(001) layer

Figure 8